\documentclass[epj]{svjour}
\usepackage{graphics}

\begin{document}

\title{Statistical field theories deformed within different calculi}
\author{A.I. Olemskoi\inst{1,2} \and S.S. Borysov\inst{2} \and I.A. Shuda\inst{2}
}                     

\institute{Institute of Applied Physics, Nat. Acad. Sci. of Ukraine, 58
Petropavlovskaya St., 40030 Sumy, Ukraine \and Sumy State University, 2
Rimskii-Korsakov St., 40007 Sumy, Ukraine}

\date{Received: date / Revised version: date}
%
\abstract{Within framework of basic-deformed and finite-difference calculi, as
well as deformation procedures proposed by Tsallis, Abe, and Kaniadakis to be
generalized by Naudts, we develop field-theoretical schemes of statistically
distributed fields. We construct a set of generating functionals and find their
connection with corresponding correlators for basic-deformed,
finite-difference, and Kaniadakis calculi. Moreover, we introduce pair of
additive functionals, whose expansions into deformed series yield both Green
functions and their irreducible proper vertices. We find as well formal
equations, governing by the generating functionals of systems which possess a
symmetry with respect to a field variation and are subjected to an arbitrary
constrain. Finally, we generalize field-theoretical schemes inherent in
concrete calculi in the Naudts spirit. From the physical point of view, we
study dependences of both one-site partition function and variance of free
fields on deformations. We show that within the basic-deformed statistics
dependence of the specific partition function on deformation has in logarithmic
axes symmetrical form with respect to maximum related to deformation absence;
in case of the finite-difference statistics, the partition function takes
non-deformed value; for the Kaniadakis statistics, curves of related
dependences have convex symmetrical form at small curvatures of the effective
action and concave form at large ones. We demonstrate that the only moment of
the second order of free fields takes non-zero values to be proportional to
inverse curvature of effective action. In dependence of the deformation
parameter, the free field variance has linearly arising form for the
basic-deformed distribution and increases non-linearly rapidly in case of the
finite-difference statistics; for more complicated case of the Kaniadakis
distribution, related dependence has double-well form.
 }
\maketitle

\section{Introduction}\label{Sec.1}

In the course of the complex system investigations, vast variety of statistical
theories has been developed \cite{4,5,TMP,6,7,Naudts,8,9,10,BasicDef}. The
approaches related are based on the principle peculiarity of the statistical
behavior of complex systems which is known to be their complicated dynamics
being spanned in the fractal phase space governed by long-range interaction or
long-time memory effects \cite{1,2,3,T}. These complications have been overcome
within framework of the standard statistical approach \cite{Callen} by means of
modification of the Boltzmann-Gibbs distribution due to deformations of the
exponential function. Formally, the approaches proposed are based on using the
extremum principle for a deformed entropic forms where the ordinary logarithm
function is substituted with some versions modified according to Tsallis
\cite{4}, Abe \cite{6}, Kaniadakis \cite{7}, Naudts \cite{Naudts}, basic
deformation procedure \cite{BasicDef}, et cetera. To our knowledge, the only
effort along a direct using the field-theoretical method for development of a
statistical scheme has been attempted in the work \cite{OSh} for the Tsallis
thermostatistics. The present article is undertaken with the purpose to
generalize the standard statistical field theory \cite{Par,Zinn} within
different versions of calculi elaborated to this moment.

Historically, the first example of such a calculus gives the basic-deformed
calculus ($q$-calculus, in other words) which has been originally introduced by
Heine and Jackson \cite{15,16} in the study of the basic hypergeometric series
\cite{17,18}. It appears the $q$-calculus does not only play a central role in
the the quantum groups and algebra belonging to mathematical branches, but have
a deep physical meaning \cite{19,20}. In this context, the studying
$q$-deformed bosons and fermions shows that thermodynamics can be built on the
formalism of $q$-calculus where the ordinary derivative is substituted by the
use of an appropriate Jackson derivative and $q$-integral \cite{21}. Moreover,
being based on a scale transformation related to the Jackson
$q$-deri\-va\-ti\-ve and $q$-integral, the basic-deformed calculus is very well
suited to describe multifractal sets \cite{22,23}. Displaying critical
phenomena of the type of growth processes, rupture, earthquake, financial
crashes, these systems reveal a discrete scale invariance with the existence of
log-periodic oscillations deriving from a partial breakdown of the continuous
scale invariance symmetry into a discrete one -- as occurs, for example, in
hierarchical lattices \cite{24,25,26,OVSh}.

Concerning the formalism on which is based our me\-thod, one needs to point out
that it reduces to using a generating functional which presents the
Fourier-Laplace transform of the partition function from the dependence on the
fluctuating distribution of an order parameter to an auxiliary field
\cite{Par,Zinn}. Due to the exponential character of this transform
determination of correlators of the order parameter is provided by
differentiation of the generating functional over auxiliary field. As was
mentioned above, fractality of the phase space of complex systems causes a
deformation of both exponential function and integral itself in the
Fourier-Laplace transform, so that the simple procedure of the differentiation
of the generating functional becomes inconsistent. Thus, main problem of our
consideration reduces to finding derivation operators, whose action keeps
form of deformed exponentials giving eigen numbers related (within the
basic-deformed calculus, such derivative reduces to the Jackson one, while
eigen number is presented by basic-number \cite{Kac}).

This paper is devoted to development of the field-theoretical scheme basing on
possible generalizations of both Fourier-Laplace transform and derivative
operator related. The work is organized as follows. In Section \ref{Sec.2}, we
yield a necessary information from the theory of both basic-deformed calculus
and finite-difference calculus ($h$-calculus); moreover, we consider as well
main peculiarities of the deformation procedures according to Tsallis, Abe,
Kaniadakis, and Naudts. Sections \ref{Sec.3}--\ref{Sec.5} are devoted to
construction of the generating functionals and finding their connection with
related correlators for basic-deformed, finite-difference, and Kaniadakis
calculi. Within the simplest harmonic approach, we find the partition functions
and the order parameter moments as functions of the deformation parameters.
Moreover, we introduce pair of additive functionals whose expansions into
deformed series yield both Green functions and their proper vertices; as well,
we find formal equations governing by the generating functionals of systems,
which possess a symmetry with respect to a field variation and are subjected to
an arbitrary constrain. In Section \ref{Sec.6}, we generalize the
field-theoretical schemes elaborated in previous Sections. Section \ref{Sec.7}
concludes our consideration and Appendix A contains calculations related to
deformed Gamma functions.

\section{Preliminaries}\label{Sec.2}

We start by referring to the standard definition of the dual pair of {\it the
basic exponentials} in form of the Taylor series \cite{Kac}
\begin{equation}
 e_q(x):=\sum_{n=0}^{\infty}\frac{x^n}{[n]_q!},\quad
 E_q(x):=\sum_{n=0}^{\infty}\frac{x^n}{[n]_q!}q^{\frac{n(n-1)}{2}}.
 \label{e}
\end{equation}
These series are seen to be determined by factorials $[n]_q!=1\cdot
2\cdot\dots\cdot[n-1]_q\cdot[n]_q$ of the basic numbers
\begin{equation}
 [n]_q:=\frac{q^n-1}{q-1}
 \label{bn}
\end{equation}
with the limit $[n]_{q\to 1}=n$ due to which the functions (\ref{e}) transform
into the ordinary exponential at $q\to1$. Since $[n]_{1/q}=[n]_q q^{-(n-1)}$,
one has the relation $[n]_{1/q}!=[n]_q!q^{-\frac{n(n-1)}{2}}$ which yields the
connection
\begin{equation}
 E_{q}(x)=e_{1/q}(x).
  \label{Ee}
\end{equation}
The mutual complementarity of the basic exponentials (\ref{e}) is apparent from
the multiplication rule \cite{Kac}
\begin{equation}
 E_{q}(x)e_{q}(y)=e_{q}(x+y)
\label{ee}
\end{equation}
according to which
\begin{equation}
E_q(x)e_q(-x)=1.
  \label{con1}
\end{equation}

The principle peculiarity of the exponentials (\ref{e}) is to keep their forms
under action of the Jackson derivative
\begin{equation}
\mathcal{D}_x^q f(x)\equiv\frac{{\rm d}_qf(x)}{{\rm d}_q x}:=\frac{f(q
x)-f(x)}{(q-1)x}.
 \label{J}
\end{equation}
Really, one has for arbitrary constants $a$ and $b$ \cite{Kac}
\begin{eqnarray}
 &\mathcal{D}_x^q e_{q}(ax+b)=a e_{q}(ax+b),\nonumber\\
 &\mathcal{D}_x^q E_{q}(ax+b)=a E_{q}(qax+b).\label{dqe}
\end{eqnarray}
In this way, for arbitrary functions $f(x)$ and $g(x)$ the Leibnitz rule reads:
\begin{eqnarray}
\mathcal{D}_x^q\left[f(x)g(x)\right]=g(qx)\mathcal{D}_x^qf(x)+f(x)\mathcal{D}_x^qg(x)\nonumber\\
=g(x)\mathcal{D}_x^qf(x)+f(qx)\mathcal{D}_x^qg(x).
\end{eqnarray}
Let us write as well the useful equations
\begin{eqnarray}
&&e_q(q^nx)=\Big(1+\left(q-1\right)x\Big)^n _qe_q(x),\nonumber\\
&&E_q(q^nx)=\frac{E_q(x)}{\Big(1-\left(q-1\right)x\Big)^n _q}
 \label{rel}
\end{eqnarray}
following from the definition (\ref{J}) and the equalities (\ref{dqe}) at $a=1$
and $b=0$. Here, we use the basic-deformed binomial \cite{Kac}
\begin{equation}
(x+y)^n _q:=(x+y)(x+qy)\dots(x+q^{n-1}y).
 \label{binom}
\end{equation}

In addition to the basic exponentials (\ref{e}) being invariant with respect to
action of the Jackson derivative (\ref{J}), its eigen-functions are known to
represent the homogeneous functions determined with the property
\begin{equation}
h(\lambda x)=\lambda^q h(x)
 \label{homfunc}
\end{equation}
where an exponent $q$ plays the role of the self-similarity degree, $\lambda$
is an arbitrary factor playing the role of deformation of self-similar systems
for which the homogeneous functions are the basis of the statistical theory
related \cite{OVSh}. To this end, the eigen-values of the Jackson derivative
(\ref{J}) determined on the set of the homogeneous functions represent the
basic numbers (\ref{bn}):
\begin{equation}
\mathcal{D}_x^\lambda h(x)=[q]_\lambda h(x),\quad
[q]_\lambda=\frac{\lambda^q-1}{\lambda-1}.
 \label{Jh}
\end{equation}

Apart from invariant action of the Jackson derivatine $\mathcal{D}_x^q$, the
exponentials (\ref{e}) are persistent as well under action of {\it the basic
integral} $\mathcal{I}_x^q$, whose property $\mathcal{D}_x^q\mathcal{I}_x^q=1$
relates to the explicit definition \cite{Kac}
\begin{equation}
 \mathcal{I}_x^q f(x)\equiv\int f(x){\rm d}_q x:=
 (1-q)x\sum_{n=0}^{\infty}f(q^n a)q^n.
  \label{int}
\end{equation}
Meaning $\mathcal{D}_x^q$ and $\mathcal{I}_x^q$ as transformation operators of
the Lee group, let us introduce the generators related:
\begin{eqnarray}
 &u_x^q:=ln_q(\mathcal{D}_x^q),\qquad j_x^q:=ln_q(\mathcal{I}_x^q);\nonumber\\
 &U_x^q:=Ln_q(\mathcal{D}_x^q),\qquad
 J_x^q:=Ln_q(\mathcal{I}_x^q).\label{UJ}
\end{eqnarray}
Here, the pair of the basic logarithmic functions $ln_q(x)$ and $Ln_q(x)$ is
used as inverse functions to the exponentials (\ref{e}). By using the
properties $u_x^q+j_x^q=0$ and $U_x^q+J_x^q=0$, one obtains the functional
expressions of the basic integral through the Jackson derivative:
\begin{eqnarray}
&\mathcal{I}_x^q=e_q\left(-u_x^q\right)=e_q\left[-ln_q\left(\mathcal{D}_x^q\right)\right],
\nonumber\\ &\mathcal{I}_x^q=E_q\left(-U_x^q\right)=
E_q\left[-Ln_q\left(\mathcal{D}_x^q\right)\right].\label{ID}
\end{eqnarray}

Let us list now main cases of deformations that complete the basic deformation
related to the exponentials (\ref{e}). Above, we have demonstrated the dual
pair of these exponentials is determined by the numbers $[n]_{q}$ and
$[n]_{1/q}$ defined by Eq. (\ref{bn}). In the case of {\it the symmetrized
$q$-calculus}, the corresponding number
\begin{equation}
 [n]_{q}:=\frac{q^n-q^{-n}}{q-q^{-1}}
 \label{sym}
\end{equation}
is obeyed the condition $[n]_{1/q}=[n]_{q}$, so that the dual exponential
$E_q(x)=e_{1/q}(x)$ coincides with the original one, $e_q(x)$. The symmetrized
$q$-calculus is a basis of the Abe statistics \cite{6}.

The third example gives {\it the $h$-exponential} that is specified by the
expression \cite{Kac}
\begin{equation}
 e_{h}(x):=(1+h)^{\frac{x}{h}}.
  \label{eh}
\end{equation}
This function is inverse to the $h$-logarithm
\begin{equation}
ln_h(x)=\frac{h\ln(x)}{\ln(1+h)}
  \label{lnh}
\end{equation}
with $\ln(x)$ being the ordinary logarithm function. It is easy to convince the
expression (\ref{eh}) may be presented in form of the series (\ref{e}) if the
basic numbers (\ref{bn}) are substituted by the $h$-numbers
\begin{equation}
 [n]_h:=\frac{hn}{\ln(1+h)}.
  \label{hd}
\end{equation}
The dual $h$-exponential can be defined as $E_{h}(x):=e_{-h}(x)$ to obey the
trivial rule
\begin{equation}\label{eeanti}
 e_h(x+y)=e_h(x)e_h(y)=E_{-h}(x)e_h(y)
\end{equation}
instead of Eq. (\ref{ee}). Setting the symmetry with respect to change of the
$h$ sign, we obtain the self-dual exponential $e_h(x)=E_h(x)\equiv e_{-h}(x)$,
with whose consideration we restrict ourselves further. In the limit $h\to 0$,
the $h$-calculus reduces naturally to the usual one.

The $h$-exponential (\ref{eh}) is obviously invariant with respect to action of
the $h$-derivative \cite{Kac}
\begin{equation}
 D_x^h f(x):=\frac{f(x+h)-f(x)}{h}.
  \label{dh}
\end{equation}
However, the properties (\ref{dqe}) take the form
\begin{eqnarray}
 && D_x^h e_h(ax+b)=d_h(a) e_h(ax+b),\nonumber \\
 && d_h(a)\equiv\frac{e_h(ah)-1}{h}=\frac{(1+h)^a-1}{h}
  \label{dhe}
\end{eqnarray}
complicated with the factor $d_h(a)$. Although this factor has the ordinary
limit $d_{h\to 0}(a)\to a$, action of the $h$-derivative (\ref{dh}) on the
exponential (\ref{eh}) does not obey a condition of the type (\ref{dqe}) for
arbitrary values of a constant $a$. To restore such a condition we shall use
the definition of the $h$-derivative
\begin{equation}
 \mathcal{D}_x^h:=[1]_h\partial_x,\quad [1]_h=\frac{h}{\ln(1+h)},\
 \partial_x\equiv\frac{\partial}{\partial x}
  \label{dhh}
\end{equation}
instead of Eq. (\ref{dh}). Being applied to the exponential (\ref{eh}), this
derivative ensures obviously the first propertiy (\ref{dqe}) with index $q$
substituted by $h$. Respectively, the $h$-integral, being inverse to the
derivative (\ref{dhh}), is defined as
\begin{equation}
\mathcal{I}_x^h f(x)=\int f(x){\rm d}_h x,\quad{\rm d}_h x=[1]_h^{-1}{\rm d}x.
 \label{ihh}
\end{equation}

Quite different example represents {\it the Tsallis exponential}
\begin{equation}
 \exp_{q}(x):=\left[1+(1-q)x\right]^{\frac{1}{1-q}}
  \label{eT}
\end{equation}
characterized by the deformed number
\begin{equation}
 [n]_{q}=\frac{n}{1+(1-q)(n-1)}.
 \label{nT}
\end{equation}
Here, the dual number $[n]_{1/q}$ relates to the exponential
\begin{equation}
{\rm Exp}_{q}(x):=\exp_{1/q}(x)=\left[\exp_q(-x/q)\right]^{-q},
  \label{deT}
\end{equation}
whose inverse value is proportional to the escort probability being the basis
of the Tsallis thermostatistics \cite{TMP,T}. It is principally important for
our aims the following: i) the Tsallis exponential (\ref{eT}) is not invariant
with respect to the Jackson derivative (\ref{J}), whereas action of the
ordinary derivative gives $\frac{\rm d}{{\rm d}x}\exp_{q}(x)=\exp_{q}^q(x)$;
ii) the dual exponential (\ref{deT}) is obeyed the condition ${\rm
Exp}^{1/q}_{q}(-qx)\exp_q(x)=1$ that does not coincide with the condition
(\ref{con1}). In the limit $q\to 1$, the Tsallis calculus reduces to the
ordinary one, while it coincides with the $h$-calculus at arbitrary values of
the deformation parameter $q=1-h/x$. It is worthwhile to stress, however, the
Tsallis exponential increases with the $x$-growth according to the power law
(\ref{eT}), while the $h$-exponential (\ref{eh}) varies exponentially. The
statistical field theory based on the Tsallis calculus has been developed in
work \cite{OSh}.

As the following example, we consider the case of {\it the Kaniadakis
deformation} when exponential and logarithm functions are defined as follows
\cite{7}:
\begin{equation}
\exp_\kappa(x):=\left[\kappa x+\sqrt{1+(\kappa x)^2}\right]^{1/\kappa},
  \label{Ke}
\end{equation}
\begin{equation}
\ln_\kappa(x):=\frac{x^\kappa-x^{-\kappa}}{2\kappa}.
  \label{Kl}
\end{equation}
Here, the deformation parameter $\kappa$ belongs to the interval $(-1,1)$ and
the limit $\kappa\to 0$ relates to the ordinary functions $\exp(x)$ and
$\ln(x)$. The exponential (\ref{Ke}) is self-dual function in the the sense
that it is obeyed the condition $\exp_\kappa(x)\exp_\kappa(-x)=1$ of the
type (\ref{con1}). However, the multiplication rule (\ref{ee}) takes the form
\cite{Ks}
\begin{equation}
\exp_\kappa(x)\exp_\kappa(y)=\exp_\kappa\left(x\stackrel{\kappa}{\oplus}y\right)
  \label{Kee}
\end{equation}
where the sum is deformed as
\begin{equation}
x\stackrel{\kappa}{\oplus}y:=x\sqrt{1+(\kappa y)^2}+y\sqrt{1+(\kappa x)^2}.
  \label{Keed}
\end{equation}
Remarkably, the rule of the type (\ref{Kee}) takes place also for the Tsallis
$q$-calculus where the deformed sum (\ref{Keed}) is written as \cite{Borges}
\begin{equation}
x\oplus_q y:=x+y+(1-q)xy.
  \label{Teed}
\end{equation}

Unfortunately, the exponential (\ref{Ke}) may not be presented in the form of
any deformed series (\ref{e}) with a number of the type (\ref{bn}). However, it
is easy to convince that this exponential is invariant with respect to action
of the derivation operator \cite{Ks}
\begin{equation}
\mathcal{D}_x^\kappa\equiv\frac{\partial}{\partial_\kappa x}:=\sqrt{1+(\kappa
x)^2}~\frac{\partial}{\partial x}.
  \label{DK}
\end{equation}
Moreover, for arbitrary constant $a$ one obtains
\begin{equation}
\mathcal{D}_x^{a\kappa}\exp_\kappa(ax)=a\exp_\kappa(ax)
  \label{dgK}
\end{equation}
instead of Eqs. (\ref{dqe}) and (\ref{dhe}). The integration operator being
inverse to the derivative (\ref{DK}) is defined in the relativistic form
\cite{Ks}
\begin{equation}
\mathcal{I}_x^\kappa f(x):=\int f(x){\rm d}_\kappa x,\quad {\rm d}_\kappa
x\equiv\frac{{\rm d}x}{\sqrt{1+(\kappa x)^2}}.
  \label{IK}
\end{equation}
The derivative property (\ref{dgK}) accompanied with the integral definition
(\ref{IK}) will be shown to be formal statement for development of the
field-theoretical scheme based on the Kaniadakis calculus.

It is worthwhile to stress the logarithm (\ref{Kl}) can be generalized to the
form \cite{Kan}
\begin{equation}
\ln_{\kappa\tau\varsigma}(x):= \frac{x^\tau\left[(\varsigma x)^\kappa
-(\varsigma x)^{-\kappa}\right]-\left(\varsigma^\kappa
-\varsigma^{-\kappa}\right)}{(\kappa+\tau)\varsigma^{\kappa}+(\kappa-\tau)\varsigma^{-\kappa}}
  \label{Klg}
\end{equation}
being solution of the functional equation
\begin{equation}
\partial_x\left[x\Lambda(x)\right]=\lambda\Lambda\left(x/\alpha\right)+\eta
  \label{Kls}
\end{equation}
with parameters
\begin{eqnarray}  \label{Klp}
&\alpha=\left(\frac{1+\tau-\kappa}{1+\tau+\kappa}\right)^{\frac{1}{2\kappa}},\nonumber\\
&\lambda=\frac{\left(1+\tau-\kappa\right)^{\frac{\tau+\kappa}{2\kappa}}}
{\left(1+\tau+\kappa\right)^{\frac{\tau-\kappa}{2\kappa}}},\\
&\eta=(\lambda-1)\frac{\varsigma^\kappa
-\varsigma^{-\kappa}}{(\kappa+\tau)\varsigma^{\kappa}+(\kappa-\tau)\varsigma^{-\kappa}}.\nonumber
\end{eqnarray}
Remarkably, the generalized logarithm (\ref{Klg}) yields known cases of the
following deformations: i) the choice of parameters $\tau=0$, $\varsigma=1$,
and $\kappa\in(-1,1)$ relates to the Kaniadakis deformation \cite{7}; ii)
$\kappa=-\tau=(1-q)/2$ -- the Tsallis deformation with parameter $q$ \cite{4};
iii) $\kappa=(q-1/q)/2$, $\tau=(q+1/q)/2$, and $\varsigma=1$ -- the Abe
deformation with parameter $q$ \cite{6}; iv) the choice $\varsigma=1$ relates
to the two-parameter logarithm proposed by Mittal, Sharma, and Taneja
\cite{M,ST}; v) the case $\tau=0$ relates to the scaled two-parameter logarithm
proposed by Kaniadakis \cite{Kan}.

Finally, we note one more example of generalized logarithms -- the functionally
deformed logarithm \cite{Naudts}
\begin{equation}
{\rm ln}_\phi(x):=\int\limits_1^x\frac{{\rm d}x'}{\phi(x')}
 \label{lnn}
\end{equation}
specified with a function $\phi(x)$. As usually, corresponding exponential
function ${\rm e}_\phi(x)$ is defined by the condition \break ${\rm
e}_\phi\left[{\rm ln}_\phi(x)\right]=x$. Formally, we may as well define a
derivation operator ${\rm D}_x^\phi$ which keeps the form of the functionally
deformed exponential according to the equation ${\rm D}_x^\phi{\rm
e}_\phi(x)=\eta_\phi{\rm e}_\phi(x)$ with an eigen-value $\eta_\phi$ fixed by
the $\phi(x)$ function choice.

Above considered examples show that a generalized calculus can be built as a
result of the following steps:
\begin{enumerate}
 \item Choose a deformed exponential ${\rm e}_\lambda(x)$ and find its dual
form ${\rm E}_\lambda(x)$ to be obeyed the multiplication rule
\begin{equation}
{\rm E}_\lambda(x){\rm e}_\lambda(y)={\rm e}_\lambda(x+y).
  \label{dual}
\end{equation}
If above exponentials may be expanded into the Taylor series of the type
(\ref{e}), their choice is fixed by numbers $[n]_\lambda$ generalizing the
expressions (\ref{bn}), (\ref{sym}), and (\ref{hd}) with parameter $q$ being
substituted with a deformation $\lambda$. In the case of the type of both
Tsallis and Kaniadakis deformations, more convenient to use a self-dual
exponential obeying the multiplication rule
\begin{equation}
{\rm e}_\lambda(x){\rm e}_\lambda(y)={\rm
e}_\lambda\left(x\stackrel{\lambda}{\oplus}y\right)
  \label{def}
\end{equation}
defined by specifying deformed sum $x\stackrel{\lambda}{\oplus}y$ [see, for
example, Eqs. (\ref{Keed}) and (\ref{Teed})].
 \item Define a deformed differentiation operator ${\rm D}^\lambda_x$ type of the
Jackson derivative (\ref{J}) according to the condition that this operator
keeps invariant forms of the generalized exponentials ${\rm e}_\lambda(x)$ and
${\rm E}_\lambda(x)$.
 \item Introduce a deformed integration operator according to the definitions
\begin{eqnarray}
{\rm I}_x^\lambda={\rm e}_\lambda\left[-\ln_\lambda\left({\rm
D}_x^\lambda\right)\right]= {\rm E}_\lambda\left[-{\rm Ln}_\lambda\left({\rm
D}_x^\lambda\right)\right] \label{IDgen}
\end{eqnarray}
generalizing Eqs. (\ref{ID}) (here, deformed logarithmic functions
$\ln_\lambda(x)$ and ${\rm Ln}_\lambda(x)$ are defined to be inverse to the
generalized exponentials related).
\end{enumerate}

As a result, we achieve the position to develop a statistical field theory that
is based on the use of a generating functional being a generalization of the
characteristic function \cite{Par,Zinn}. This function is known to be presented
by the Fourier-Laplace transform
\begin{equation}
p(j):={\rm I}_x^\lambda\left[p(x){\rm E}_\lambda(jx)\right]=\int p(x){\rm
E}_\lambda (jx){\rm d}_\lambda x
  \label{cf}
\end{equation}
of the probability distribution $p(x)$. The key point is that deformed
exponential standing within integrand of the characteristic function (\ref{cf})
is eigen-function of the deformed derivative operator ${\rm D}_j^\lambda$ with
eigen value $d_\lambda(x)$. As a result, multiple differentiation of this
function over auxiliary variable $j$ keeps its exponential form to yield the
moments
\begin{eqnarray}
\left<\left[d_\lambda(x)\right]^n\right>_\lambda &&
:=\int\left[d_\lambda(x)\right]^n p(x){\rm d}_\lambda x\nonumber\\ &&
=\left.\left({\rm D}_j^\lambda\right)^np(j)\right|_{j=0}
\end{eqnarray}
of an order parameter $d_\lambda(x)$.

\section{Basic-deformed statistics}\label{Sec.3}

Let us consider a statistical system, whose distribution over states ${\bf
x}=\{{\bf r}_a,{\bf p}_a\}$ in the phase space of particles $a=1,\dots,N$,
$N\to\infty$ with coordinates ${\bf r}_a$ and momenta ${\bf p}_a$ is determined
by a Hamiltonian $H=H\left({\bf x}\right)$. We are interested in study of the
coarse space distribution $\phi({\bf r})$ of an order parameter $\phi$. Within
the coarse grain approximation, thermostatistics of the deformed system is
governed by the partition functional
\begin{eqnarray}
 \mathcal{Z}_q\{\phi\}&&:=\int e_q\left[-\beta H\left({\bf x}\right)\right]
 \delta\left[\phi-\phi\left({\bf x}\right)\right]{\rm d}_q{\bf x}\nonumber\\
 &&\equiv
 e_q\left(-S\{\phi\}\right).
 \label{Z}
\end{eqnarray}
Here, ${\rm d}_q{\bf x}=(q-1){\bf x}$ stands for the basic-deformed
differential, $S=S\{\phi\}$ is an effective action, and one takes into account
also that thermostatistical distribution of a basic-deformed system is
proportional to the $q$-deformed exponential $e_q\left[-\beta H\left({\bf
x}\right)\right]$ with the inverse temperature $\beta$ measured in the energy
units \cite{BasicDef}.

The principle peculiarity of the definition (\ref{Z}) is that both
thermostatistical exponential and integral over the phase space are the basic
deformed ones. To this end, we need to use the basic-deformed Laplace transform
\begin{eqnarray}
 \mathcal{Z}_{q}\{J\}&&:=\int\mathcal{Z}_q\{\phi\}E_{q}\{J\cdot\phi\}\{{\rm d}_q\phi\}\nonumber\\
 &&=\int e_{q}\left(-S\{\phi\}+J\cdot\phi\right)\{{\rm d}_q\phi\}
 \label{Z(J)}
\end{eqnarray}
where the last equation is written with accounting the property (\ref{ee}). For
the sake of simplicity, we use the lattice representation to describe the
coordinate dependence by means of the index $i=1,\dots,N$ in the
shorthands $J\cdot\phi\equiv\sum_i J_i\phi_i$ and $\{{\rm
d}_q\phi\}\equiv\prod_i{\rm d}_q\phi_i$.

According to the rules (\ref{dqe}) the $n$-fold differentiation of the last
expression for the generating functional (\ref{Z(J)}) yields
\begin{eqnarray}
&&\left(\mathcal{D}_{J_1}^q\dots\mathcal{D}_{J_n}^q\right)\mathcal{Z}_q\{J\}\nonumber\\
&&=\int\left(\phi_{i_1}\dots\phi_{i_n}\right)
e_{q}\left(-S\{\phi\}+J\cdot\phi\right)\{{\rm d}_q\phi\}.\ \ \
 \label{DDD}
\end{eqnarray}
The right hand site of this equality determines the correlator
\begin{equation}
\left<\phi_{i_1}\dots\phi_{i_n}\right>_q:=\mathcal{Z}_q^{-1}\int\left(\phi_{i_1}\dots\phi_{i_n}\right)
\mathcal{Z}_q\{\phi\}\{{\rm d}_q\phi\}
 \label{C}
\end{equation}
where the coefficient is inversely proportional to the partition function
\begin{equation}
 \mathcal{Z}_{q}:=\int\mathcal{Z}_q\{\phi\}\{{\rm d}_q\phi\}
 =\int e_{q}\left(-S\{\phi\}\right)\{{\rm d}_q\phi\}.
 \label{ZZ}
\end{equation}
Combination of the last equalities allows for one to express an arbitrary
correlator through the basic-deformed derivatives of the generating functional
(\ref{Z(J)}):
\begin{eqnarray}
&&\left<\phi_{i_1}\dots\phi_{i_n}\right>_q\nonumber\\ &&=
\mathcal{Z}_q^{-1}\left.\left(\mathcal{D}_{J_1}^q\dots\mathcal{D}_{J_n}^q\right)
\mathcal{Z}_q\{J\}\right|_{J_1,\dots,J_n=0}.
 \label{CC}
\end{eqnarray}

\begin{figure}
 \resizebox{0.5\textwidth}{!}{
 \includegraphics{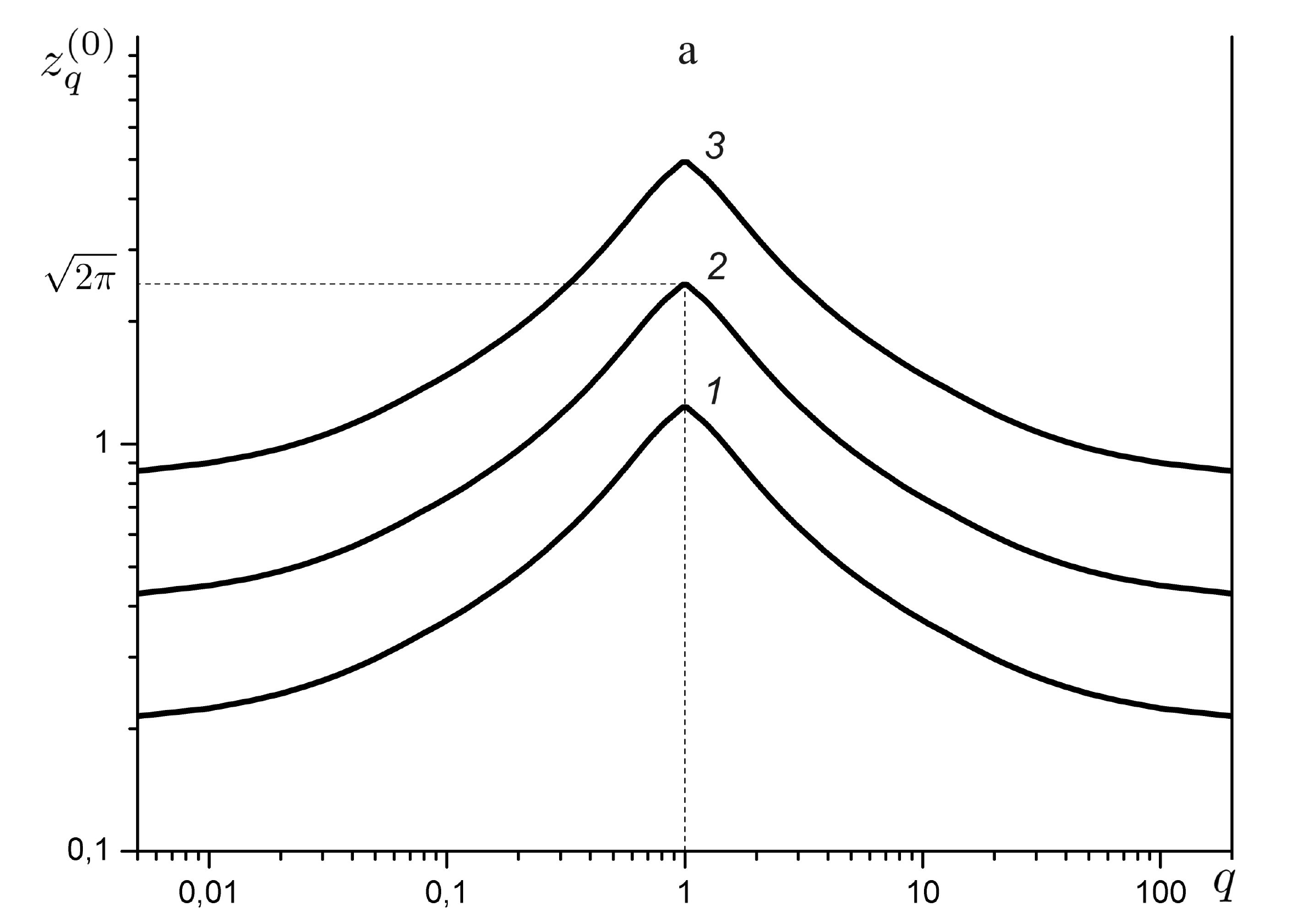}}\\
 \resizebox{0.5\textwidth}{!}{
 \includegraphics{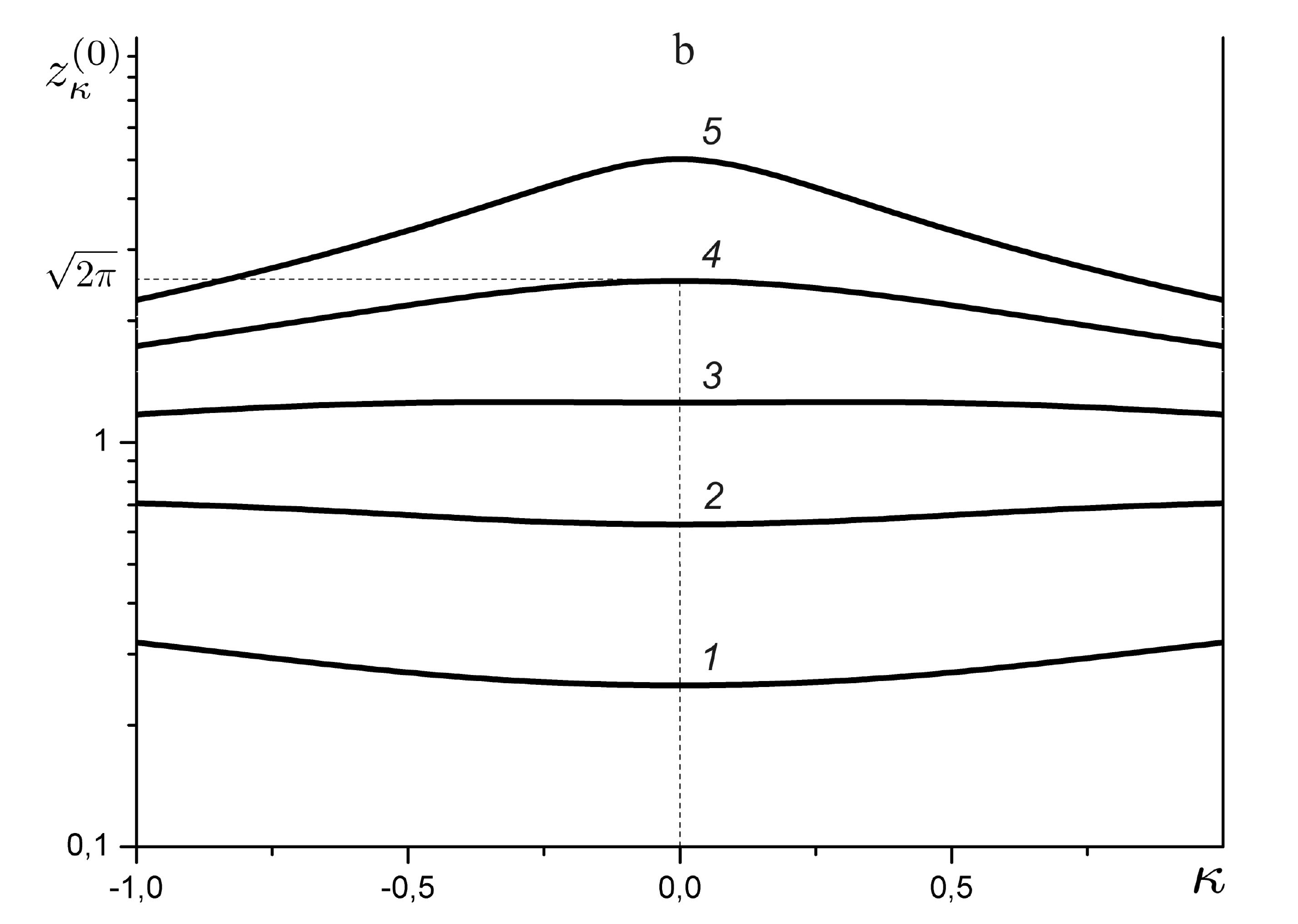}}\\
  \caption{Dependences of the one-site partition functions on the deformation
parameters: a) within the basic-deformed statistics at $\Delta=0.5,1,2$ (curves
1, 2, 3, respectively); b) within the Kaniadakis statistics (curves 1--5
correspond to $\Delta=0.1, 0.25, 0.5, 1.0, 2.0$).} \label{z0q}
\end{figure}

Within framework of the harmonic approach, effective action takes the deformed
parabolic form
\begin{equation}
S^{(0)}\{\phi\}=\sum_i^N S^{(0)}\{\phi_i\},\quad
S^{(0)}\{\phi_i\}\equiv\frac{(\phi_i)_q^2 }{[2]_q\Delta^2}
 \label{S0}
\end{equation}
where the deformed binom (\ref{binom}) is used and $\Delta^2$ stands for the
inverse curvature. To apply the rule (\ref{ee}) for the basic-deformed
exponential in the generating functional (\ref{Z(J)}) let us suppose the
symmetry with respect to substituting the deformation parameter $q$ by the
inverse value $1/q$. Then, it is convenient to separate whole lattice into odd
sites $i'$ and even ones $i{''}$ and use Eq. (\ref{ee}) for each of couples
$i',i{''}$. To this end, the exponential in the partition function (\ref{ZZ})
is transformed as follows:
\begin{eqnarray}
&& e_{q}\left(-S^{(0)}\{\phi\}\right)=e_{q}\left(\sum_i^N
S^{(0)}\{\phi_i\}\right)\nonumber\\ &&=
\prod_{i'}^{[N/2]}e_{q}\left(S^{(0)}\{\phi_{i'}\}\right)~
\prod_{i{''}}^{[N/2]}e_{1/q}\left(S^{(0)}\{\phi_{i{''}}\}\right)
 \label{eee}
\end{eqnarray}
where square brackets denote the integer of the fraction
$N/2$.\footnote{Generally speaking, a statistical ensemble might comprise of
odd number of particles $N$ that must arrive at unmatched factor in products
(\ref{eee}). However, this factor is negligible within the thermodynamic limit
$N\to\infty$.} As a result, the generating functional (\ref{Z(J)}) takes the
multiplicative form
\begin{equation}
\mathcal{Z}_q^{(0)}\{J\}=\prod_{i'}^{[N/2]}z_q^{(0)}(J_{i'})\prod_{i{''}}^{[N/2]}
z_{1/q}^{(0)}(J_{i{''}}).
 \label{Zz}
\end{equation}
As show simple calculations in Appendix A, each of multipliers related to one
site is determined by the expression
\begin{equation}
z_q^{(0)}(J)=\frac{2\Delta}{\sqrt{[2]_q}}\gamma_q\left(\frac{1}{2}\right)
E_q\left[\frac{q}{[2]_q}(\Delta J)^2\right]
 \label{z0}
\end{equation}
where the basic-deformed $\gamma$-function is defined by the first equation
(\ref{gamma}). Respectively, specific partition function $z_q^{(0)}\equiv
z_q^{(0)}(J=0)$ reads:
\begin{equation}
z_q^{(0)}=\frac{2\Delta}{\sqrt{[2]_q}}\gamma_q\left(\frac{1}{2}\right),\quad
[2]_q=1+q.
 \label{G0}
\end{equation}
As shows Figure \ref{z0q}a, the dependence of this function on the deformation
parameter has in logarithmic axes symmetrical form with respect to the maximum
point $q=1$.

According to the definition (\ref{CC}), within the harmonic approach the order
parameter is determined as
\begin{eqnarray}
&&\left<\phi\right>_q^{(0)}=\nonumber\\
&&\left.\frac{q}{[2]_q}\frac{(1+q)-q^2(q-1)\frac{\Delta^2J^2}{[2]_q}}
{\Big(1-q(q-1)\frac{\Delta^2J^2}{[2]_q}\Big)^2
_q}E_q\left[\frac{q}{[2]_q}(\Delta J)^2\right]\Delta^2 J
\right|_{J=0}=0\nonumber
\end{eqnarray}
where the second relation (\ref{rel}) is used. In similar manner, cumbersome
but simple calculations give the field variance
\begin{equation} \label{E11}
\left<\phi^2\right>_q^{(0)}=\Delta^2 q.
\end{equation}
Thus, the basic-deformed distribution of free fields has zero moment of the
first order and the variance, being proportional to the inverse curvature of
the related action (\ref{S0}) and dependent on the deformation parameter
linearly.

To develop a deformed perturbation theory one ought, as usually, to pick out an
unharmonic contribution $V=V\{\phi\}$ in total action $S=S^{(0)}+V$
\cite{Zinn}. Then, the basic-deformed exponential can be written as follows:
\begin{equation} \label{E2}
e_q(-S)=E_q(-V)e_q\left(-S^{(0)}\right).
\end{equation}
As a result, the formal expansion of the exponential $E_q(-V)$ in power series
with consequent use of the differentiation rules (\ref{dqe}) allows for one to
present the generating functional (\ref{Z(J)}) in the convenient form
\begin{equation} \label{PT}
\mathcal{Z}_q\{J\}=E_q\big(-V\left\{\mathcal{D}^q_J\right\}\big)
\mathcal{Z}_q^{(0)}\{J\}.
\end{equation}
Further, making use of the perturbation scheme with implementation of related
diagram technics is straightforward \cite{Zinn}. Moreover, the thermodynamic
limit $N\to\infty$ allows for one to use the Wick theorem to express higher
correlators through the variance (\ref{E11}).

Similarly to the ordinary field scheme \cite{Zinn}, an inconvenience of the
above approach is that the generating functional (\ref{Z(J)}) is non-additive
value. To escape this drawback one should introduce the Green functional
\begin{equation}
\mathcal{G}_q:=ln_q\left(\mathcal{Z}_q\right)
 \label{G}
\end{equation}
being deformed logarithm of the functional (\ref{Z(J)}). It worthwhile to note
the function of the deformed logarithm has not an explicit form to be defined
by the inverse exponential function
$\mathcal{Z}_q=e_q\left(\mathcal{G}_q\right)$ given by the first series
(\ref{e}).

\begin{figure}
 \resizebox{0.45\textwidth}{!}{
 \includegraphics{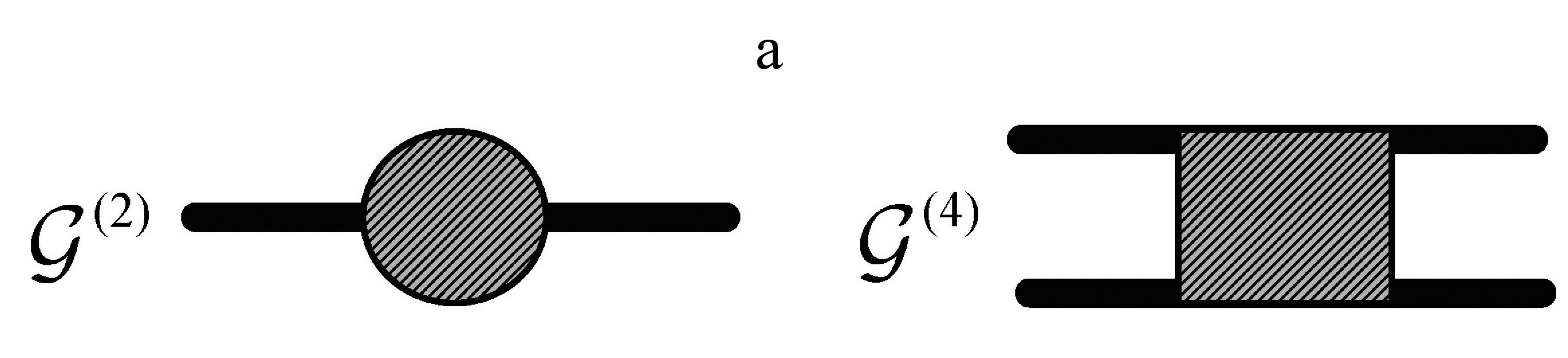}}\\
 \resizebox{0.45\textwidth}{!}{
 \includegraphics{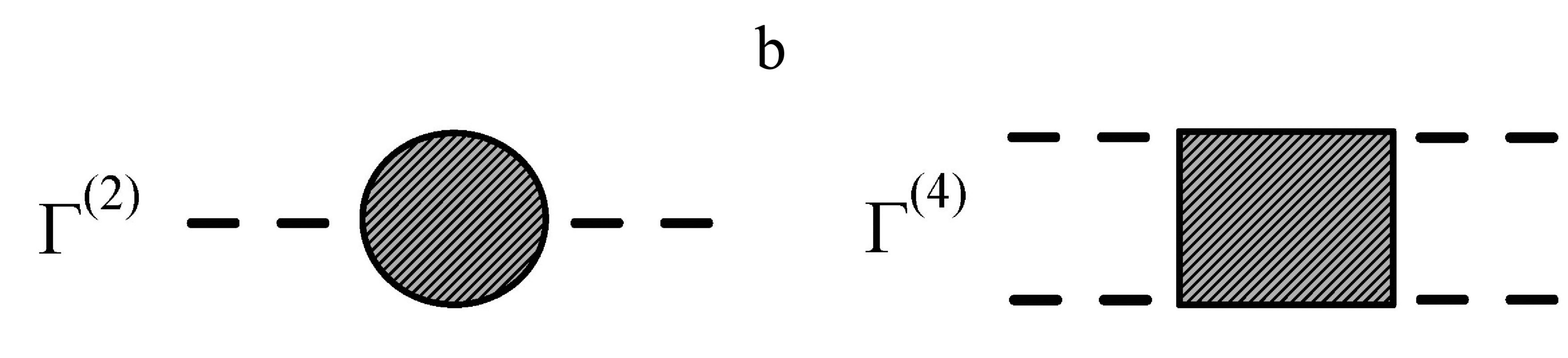}}
  \caption{Diagram representations of Green functions (a)
  and their irreducible parts (b).}\label{G.pdf}
\end{figure}

Since the Green functional (\ref{G}) depends on an auxiliary field $J$, it can
be more convenient to use a conjugate functional $\Gamma_q=\Gamma_q\{\phi\}$,
whose dependence of the initial field $\phi$ is provided by the Legendre
transformation
\begin{equation}
\Gamma_q\{\phi\}:=\sum_i J_i\phi_i-\mathcal{G}_q\{J\}.
 \label{L}
\end{equation}
The pair of the functionals $\mathcal{G}_q\{J\}$ and $\Gamma_q\{\phi\}$ plays
the role of conjugated potentials, whose basic-deformed variation yields the
state equations
\begin{equation}
\phi_i=\mathcal{D}_{J_i}^q\mathcal{G}_q\quad\Leftrightarrow\quad J_i=\mathcal
{D}_{\phi_i}^q\Gamma_q.
 \label{SE}
\end{equation}
Being analytical functional, these potentials can be presented by the following
series:
\begin{equation}
\mathcal{G}_q\{J\}=\sum\limits_{n=1}^{\infty}\frac{1}{[n]_q!}\sum_{i_1\dots
i_n}\mathcal{G}^{(n)}_{i_1\dots i_n}J_{i_1}\dots J _{i_n},
 \label{DJ}
\end{equation}
\begin{eqnarray}
\Gamma_q\{\phi\}=\sum\limits_{n=1}^{\infty}\frac{1}{[n]_q!}\sum_{i_1\dots i_n}
\Gamma^{(n)}_{i_1\dots i_n}\eta_{i_1}\dots\eta_{i_n}, \label{DS}\\
\eta_i\equiv\phi_i-\mathcal{G}^{(1)}_{i_1}.\nonumber
\end{eqnarray}
To this end, related kernels $\mathcal{G}^{(n)}_{i_1\dots i_n}$ and
$\Gamma^{(n)}_{i_1\dots i_n}$ reduce to $n$-particle Green function and its
irreducible part, respectively. Within the diagram representation, these
kernels are depicted in Figure \ref{G.pdf}.

Following the standard field scheme \cite{Zinn}, we show further that the
generating functional (\ref{Z(J)}) obeys some formal relations. The first
displays a system symmetry with respect to the basic-deformed variation in the
form
\begin{equation} \label{var}
\delta_q\phi_i=\epsilon_q f_i\{\phi\}
\end{equation}
given by an analytical functional $f_i\{\phi\}$ in the limit $\epsilon_q\to 0$.
Due to this variation the integrand of the last expression of the functional
(\ref{Z(J)}) is transformed in the following manner:
\begin{eqnarray} \label{varZ}
e_q\left[-S\{\phi+\delta_q\phi\}+J\cdot\left(\phi+\delta_q\phi\right)\right]\nonumber\\
\simeq e_q\left[\left(-S\{\phi\}+J\cdot\phi\right)+ \left(-\frac{\partial
S}{\partial\phi_i}+J_i\right)\delta_q\phi_i\right]\nonumber\\ =
e_q\Big(-S\{\phi\}+J\cdot\phi\Big)~E_q\left[\left(-\frac{\partial
S}{\partial\phi_i}+J_i\right)\delta_q\phi_i\right]\nonumber\\ \simeq
e_q\Big(-S\{\phi\}+J\cdot\phi\Big)\left[1+\left(-\frac{\partial
S}{\partial\phi_i}+J_i\right)\epsilon_q f_i\{\phi\}\right]
\end{eqnarray}
where the sum over repeated indexes is implied and the second expansion
(\ref{e}) is taken into account. On the other hand, the Jacobian determinant
appearing due to passage from $\phi$ to $\phi+\delta_q\phi$ gives the factor
$1+\left({\partial f_i}/{\partial\phi_i}\right)\epsilon_q$. As a result,
collecting multipliers which include the infinitesimal value $\epsilon_q$, one
obtains from the invariance property of the generating functional (\ref{Z(J)}):
\begin{equation} \label{EM2}
\Bigg[f_i\left\{\mathcal{D}_{J}^q\right\}\left(\frac{\partial
S}{\partial\phi_i}\left\{\mathcal{D}_{J}^q\right\}-J_i\right)-\frac{\partial
f_i}{\partial\phi_i}\left\{\mathcal{D}_{J}^q\right\}\Bigg]\mathcal{Z}_q\{J\}=0.
\end{equation}
Here, we use the first of the state equations (\ref{SE}) for operator
representations of the type
$f_i\left\{\phi\right\}e_q\left(-S\{\phi\}+J\cdot\phi\right)=
f_i\left\{\mathcal{D}_{J}^q\right\}e_q\left(-S\{\phi\}+J\cdot\phi\right)$. At
condition $f_i\left\{\phi\right\}={\rm const}$, the equation (\ref{EM2}) takes
the simplified form, following directly from the generating functional
(\ref{Z(J)}) after variation over the field $\phi$.

The second of above pointed equations allows for one to take into account an
arbitrary condition $F_j\{\phi\}=0$, $j=1,2,\dots$ for the set of fields
$\{\phi_i\}$, $i=1,2,\dots,N$ to be found. Taking into account of this
condition is achieved by inserting the $\delta$-functional $\delta\{F\}$ into
the integrand of the expression (\ref{Z(J)}) that results in introducing the
prolonged form
\begin{eqnarray} \label{I2}
&&\mathcal{Z}_q^{(F)}\{J\}\\ &&:=\int
e_q\left[-S\{\phi\}\right]E_q\left\{J\cdot\phi+\lambda\cdot F\right\}\{{\rm
d}_q\phi\}\{{\rm d}_q\lambda\}.\nonumber
\end{eqnarray}
Then, variation over an auxiliary variables $\lambda_j$, $j=1,2,\dots$ yields
the desired result
\begin{equation} \label{EM3}
F_i\left\{\mathcal{D}_{J}^q\right\}\mathcal{Z}_q^{(F)}\{J\}=0.
\end{equation}
In comparison with the standard field theory \cite{Zinn}, the principle
peculiarity of the equalities (\ref{EM2}) and (\ref{EM3}) is that they contain
the Jackson derivative $\mathcal{D}_{J}^q$ instead of the ordinary variation
$\delta/\delta J$.

\section{Finite-difference statistics}\label{Sec.4}

Within framework of the finite-difference statistics, the field scheme is based
on the definitions (\ref{eh})--(\ref{eeanti}), (\ref{dhh}) and (\ref{ihh})
inherent in the $h$-calcu\-lus \cite{Kac} to be developed in analogy with
consideration stated in the previous section. With using the
$h$-exponen\-ti\-al (\ref{eh}), the partition functional takes the form of the
$h$-integral
\begin{eqnarray} \label{Zh}
 Z_h\{\phi\}&&:=\int e_h\left[-\beta H\left({\bf x}\right)\right]
 \delta\left[\phi-\phi\left({\bf x}\right)\right]{\rm d}_h{\bf x}\nonumber\\ &&\equiv
 e_h\left(-S\{\phi\}\right)
\end{eqnarray}
instead of the expression (\ref{Z}). Respectively, the generating functional
(\ref{Z(J)}) is written as
\begin{eqnarray} \label{Zh(J)}
 Z_h\{J\}&&:=\int Z_h\{\phi\}e_h\{J\cdot\phi\}\{{\rm d}_h\phi\}\nonumber\\
 &&=\int e_h\left(-S\{\phi\}+J\cdot\phi\right)\{{\rm d}_h\phi\}.
\end{eqnarray}
Similarly to action of the Jackson derivative on the basic exponential, the
differentiation operator (\ref{dhh}) gives the correlator (\ref{C}) with $q$
substituted by $h$ in the following form [cf. Eq. (\ref{CC})]
\begin{eqnarray} \label{CCh}
&&\left<\phi_{i_1}\dots\phi_{i_n}\right>_h\nonumber
\\ &&= Z_h^{-1}\left.\left(\mathcal{D}_{J_1}^h\dots\mathcal{D}_{J_n}^h\right)
Z_h\{J\}\right|_{J_1,\dots,J_n=0}.
\end{eqnarray}

As in Eq. (\ref{S0}), the harmonic action
\begin{equation} \label{S0h}
S^{(0)}\{\phi\}=\sum_i^N \frac{\phi_i^2}{[2]_h\Delta^2},\quad
[2]_h=\frac{2h}{\ln(1+h)}
\end{equation}
is determined with the inverse curvature $\Delta^2$. Then, the generating
functional (\ref{Zh(J)}) is expressed by the product
\begin{equation} \label{Zzh}
\mathcal{Z}_h^{(0)}\{J\}=\prod_i^Nz_h^{(0)}(J_i)
\end{equation}
of the type (\ref{Zz}) with the one-particle factor (\ref{zJh}). Taking into
account the property (\ref{gammah}) of the $h$-gamma function, one obtains the
expression
\begin{equation}
z_0^{(0)}(J)=\sqrt{2\pi}\Delta e_h\left[\frac{[1]_h}{2}(\Delta J)^2\right]
 \label{z01}
\end{equation}
reducing to the standard form in the limit $h\to 0$. Moreover, the specific
partition function
\begin{equation}
z_h^{(0)}
=\int\limits_{-\infty}^{+\infty}e_h\left(-\frac{\phi^2}{[2]_h\Delta^2}\right){\rm
d}_h\phi
  \label{zzh}
\end{equation}
related to $J=0$ takes the non-deformed value $z_h^{(0)}=\sqrt{2\pi}\Delta$ for
arbitrary parameters $h$.

\begin{figure}[!htb]
\resizebox{0.5\textwidth}{!}{
 \includegraphics{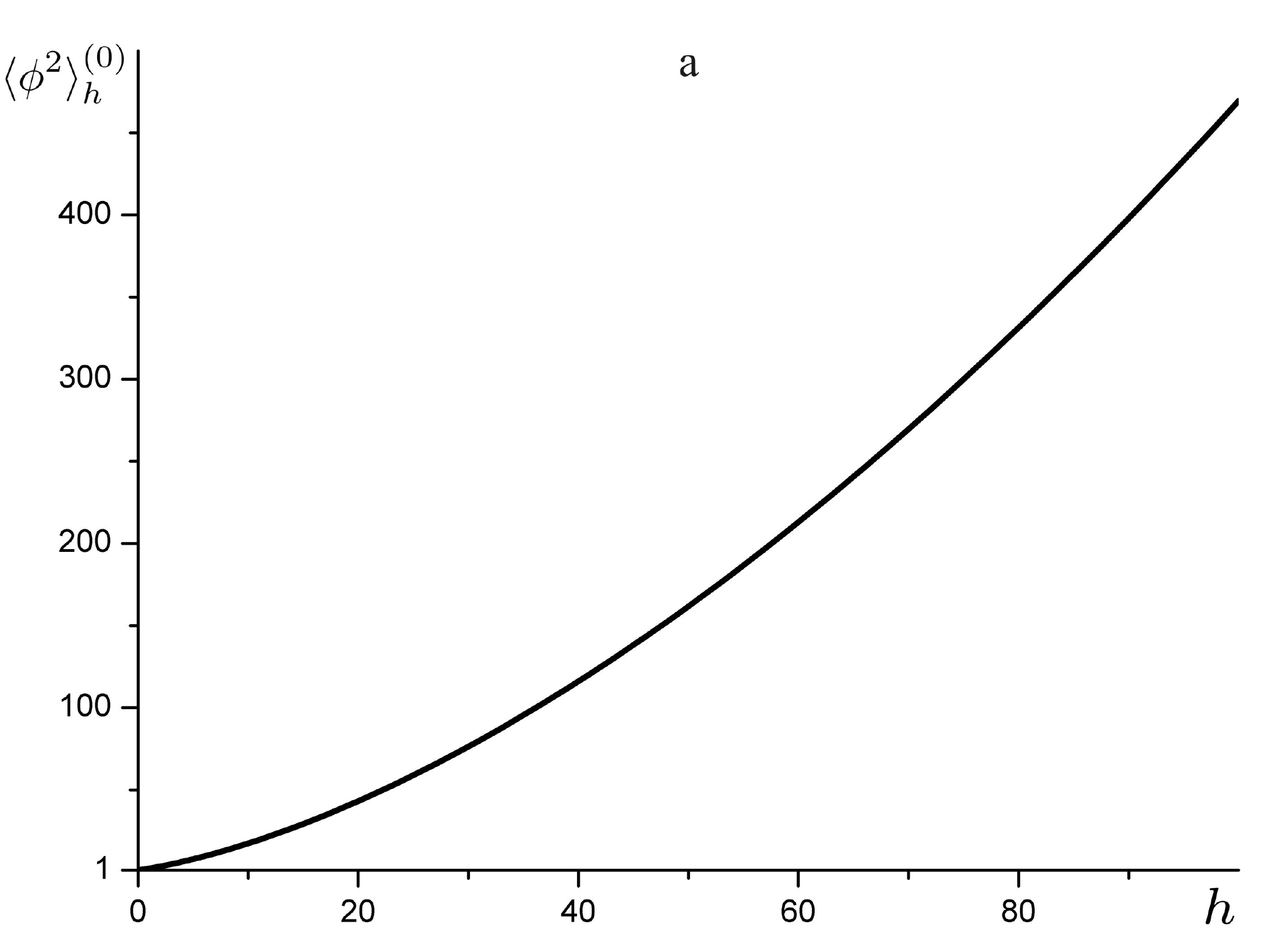}}\\
\resizebox{0.5\textwidth}{!}{
 \includegraphics{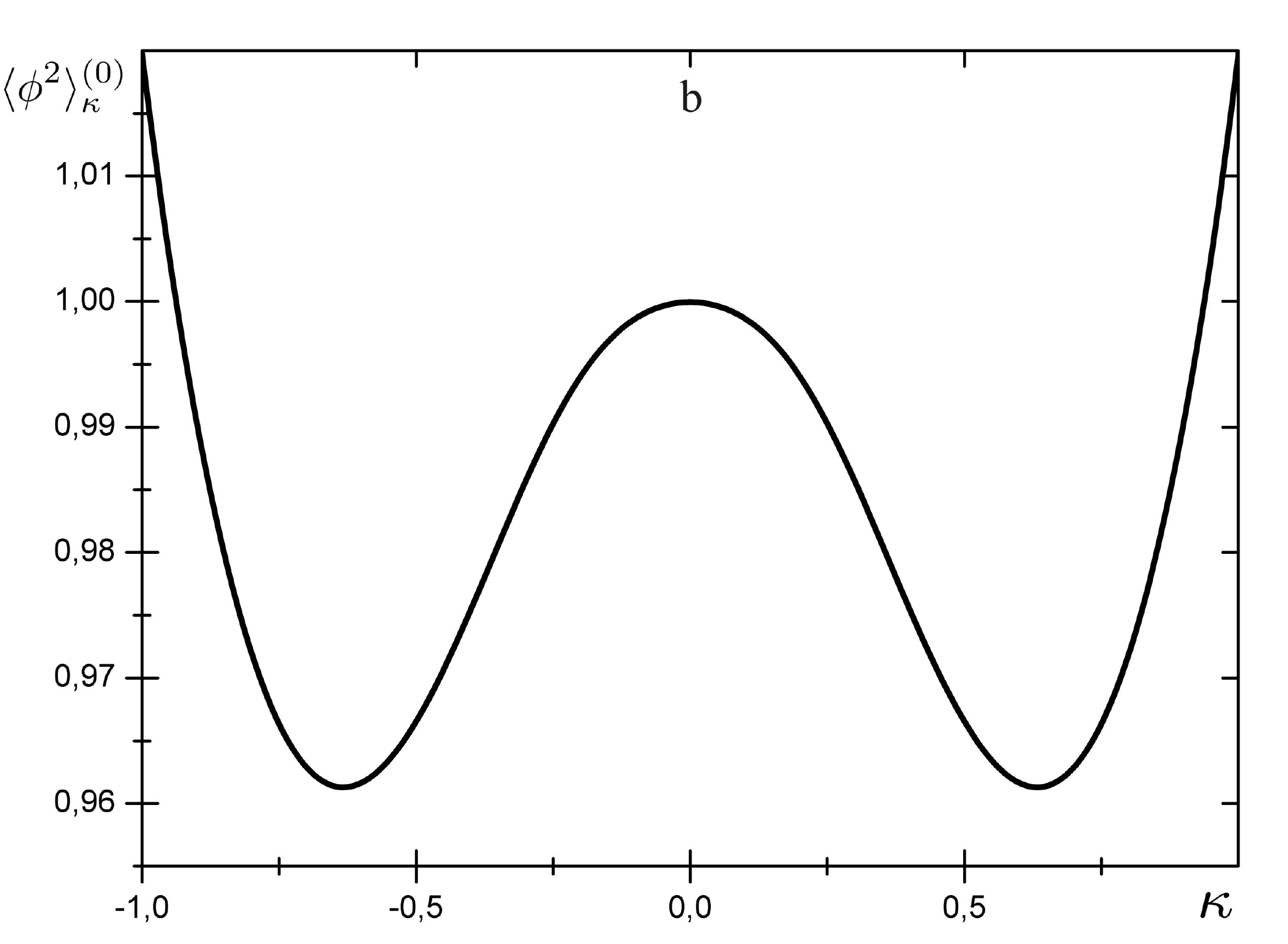}}\\
\caption{Variances of free fields versus deformation parameters within the
$h$-statistics (a) and the Kaniadakis one (b) at $\Delta=1$.} \label{z0hh}
\end{figure}

According to the definition (\ref{CCh}), the mean value of $h$-deformed free
fields equals
\begin{equation}
\left<\phi\right>_h^{(0)}=\left.[1]_he_h\left[\frac{[1]_h}{2}(\Delta
J)^2\right]\Delta^2 J\right|_{J=0}=0.
 \label{E1h}
\end{equation}
Respectively, the variance related is written as
\begin{equation}
\left<\phi^2\right>_h^{(0)}=\Delta^2[1]_h^2,\quad [1]_h=\frac{h}{\ln(1+h)}
 \label{E11h}
\end{equation}
instead of Eq. (\ref{E11}). Thus, similarly to the basic-deformed distribution
the $h$-deformed free fields have zero moment of the first order and the
variance being proportional to the inverse curvature of the action (\ref{S0h})
related. However, dependence of the variance (\ref{E11h}) on the deformation
parameter appears to be more strong than the linear dependence (\ref{E11})
related to the case of the basic deformation (see Figure \ref{z0hh}a).

The $h$-deformed perturbation theory is built in the complete accordance with
the scheme stated in the previous section, with the only difference that the
dual $q$-exponential $E_q(x)$ should be substituted with the self-dual
$h$-exponential $e_h(x)$. Along this line, making use of the symbolic
perturbation expansion of the type (\ref{E2}) yields the generating functional
\begin{equation} \label{PTh}
\mathcal{Z}_h\{J\}=e_h\left(-V\left\{\mathcal{D}^h_J\right\}\right)
\mathcal{Z}_h^{(0)}\{J\}
\end{equation}
instead of Eq. (\ref{PT}). As well, application of the diagram technics and the
Wick theorem appears to be straightforward. Moreover, following to the standard
line \cite{Zinn}, one should supplement the non-additive functional
(\ref{Zh(J)}) by the Green functional
$\mathcal{G}_h:=ln_h\left(\mathcal{Z}_h\right)$ where the $h$-logarithm
is defined as (\ref{lnh}) to be inverse to the $h$-exponential
(\ref{eh}).

The functional $\Gamma_h=\Gamma_h\{\phi\}$ conjugated to the Green functional
$\mathcal{G}_h\{J\}$ and the state equations related are defined by the
Legendre transformation (\ref{L}) and Eqs. (\ref{SE}), respectively, where
subscripts $q$ are substituted by indexes $h$. With this substitution, the
kernels of the series (\ref{DJ}) and (\ref{DS}) are reduced to the $n$-particle
Green function $\mathcal{G}_{i_1\dots i_n}^{(n)}$ and its irreducible part
$\Gamma_{i_1\dots i_n}^{(n)}$ as is depicted graphically in Figure \ref{G.pdf}.
Finally, the formal relations for the generating functional (\ref{Zh(J)}) take
the forms of equations (\ref{EM2}), (\ref{I2}), and (\ref{EM3}) to display a
system symmetry with respect to the $h$-deformed variation
$\delta_h\phi_i(x)=\phi_i(x+h)-\phi_i(x)$ and take into account an arbitrary
condition $F_j\{\phi\}=0$, $j=1,2,\dots$ for the set of fields $\{\phi_i\}$,
$i=1,2,\dots,N$ (in above pointed equations, one should again substitute $q$ by
$h$).

\section{Kaniadakis statistics}\label{Sec.5}

Taking into account close likeness between basic-deformed, Tsallis and
Kaniadakis statistics, we present the latter basing on the field-theoretical
schemes developed in Section \ref{Sec.3} and Ref. \cite{OSh}. In so doing, one
needs to substitute the Tsallis deformed exponentials (\ref{eT}) by the
Kaniadakis ones (\ref{Ke}) and take into account the multiplication rule
(\ref{Kee}) defined with the deformed sum (\ref{Keed}). As a result, the
generating functional (\ref{Z(J)}) takes the form
\begin{eqnarray}
 \mathcal{Z}_{\kappa}\{J\}&&:=\int\mathcal{Z}_\kappa\{\phi\}\exp_{\kappa}\{J\cdot\phi\}\{{\rm d}_\kappa\phi\}\nonumber\\
 &&=\int\exp_{\kappa}\left(-S\{\phi\}\stackrel{\kappa}{\oplus}J\cdot\phi\right)\{{\rm
 d}_\kappa\phi\}.
 \label{ZK(J)}
\end{eqnarray}
According to the property (\ref{dgK}), the correlator [cf. with Eq. (\ref{CC})]
\begin{eqnarray}
&&\left<\phi_{i_1}\dots\phi_{i_n}\right>_\kappa\nonumber\\ &&=
\mathcal{Z}^{-1}_\kappa\left.\left(\mathcal{D}_{J_1}^{\kappa\phi_1}\dots\mathcal{D}_{J_n}^{\kappa\phi_n}\right)
\mathcal{Z}_\kappa\{J\}\right|_{J_1,\dots,J_n=0}
 \label{CCK}
\end{eqnarray}
is determined by the partition function
$\mathcal{Z}_{\kappa}=\mathcal{Z}_{\kappa}\{J=0\}$ and the Kaniadakis
derivative (\ref{DK}). Then, with using the harmonic action
\begin{equation}
S^{(0)}\{\phi\}=\sum_i^N\frac{\phi_i^2}{2\Delta^2}
 \label{S0K}
\end{equation}
the generating functional (\ref{ZK(J)}) reduces to the product
\begin{equation}
\mathcal{Z}_\kappa^{(0)}\{J\}=\prod_{i}^{N}z_\kappa^{(0)}(J_{i})
 \label{ZzK}
\end{equation}
of the one-site factors
\begin{equation}
z_\kappa^{(0)}(J)=\int\limits_{-\infty}^\infty\exp_\kappa\left(-\frac{\phi^2}
{2\Delta^2}\right)\exp_\kappa\left(J\phi\right){\rm d}_\kappa\phi.
 \label{zK}
\end{equation}
The specific partition function $z_{\kappa}^{(0)}=z_{\kappa}^{(0)}(J=0)$ takes
the explicit form
\begin{equation}
z_{\kappa}^{(0)}=
\int\limits_{-\infty}^\infty\left[\sqrt{1+\left(\frac{\kappa\phi^2}{2\Delta^2}\right)^2}
-\frac{\kappa\phi^2}{2\Delta^2}\right]^{\frac{1}{\kappa}}\frac{{\rm
d}\phi}{\sqrt{1+\left(\kappa\phi\right)^2}}.
 \label{VK}
\end{equation}
According to Eqs. (\ref{CCK}) and (\ref{zK}), the first moment is
$\left<\phi\right>_{\kappa}^{(0)}=0$, while the definition of the free field
variance $\left<\phi^2\right>_{\kappa}^{(0)}$ is achieved by inserting the
$\phi^2$ factor into the integrand of the integral (\ref{VK}) and dividing by
$z_{\kappa}^{(0)}$.

As shows Figure \ref{z0q}b, dependence of the one-site partition function
(\ref{VK}) on the deformation parameter $\kappa$ has a symmetrical form with
respect to the point $\kappa=0$. Characteristically, the $|\kappa|$ arising
results in the $z_{\kappa}^{(0)}$ increase at small values $\Delta^2$ of the
inverse curvature of action (\ref{S0K}), while the $\Delta$ growth transforms
the concave curve of the dependence $z_{\kappa}^{(0)}(\kappa)$ into the convex
one. What about the dependence $\left<\phi^2\right>_{\kappa}^{(0)}(\kappa)$ for
the variance of the $\kappa$-deformed free fields, Figure \ref{z0hh}b
visualizes more complicated curve: first, the $|\kappa|$ growing results in the
$\left<\phi^2\right>_{\kappa}^{(0)}$ decrease from the value
$\left<\phi^2\right>_{\kappa=0}^{(0)}=\Delta^2$, after that the field variance
increases before an anomalous value
$\left<\phi^2\right>_{|\kappa|=1}^{(0)}>\Delta^2$.

\section{Generally deformed statistics}\label{Sec.6}

The examples considered in Sections \ref{Sec.3}--\ref{Sec.5} and Ref.
\cite{OSh} show the generalization of the deformed statistics should be
performed along the two different lines. The first generalizes the
basic-deformed and $h$-statistics, the second makes the same for statistics of
the type proposed by Tsallis and Kaniadakis. Such a partition is stipulated by
the principle difference between the multiplication rules (\ref{ee}) and
(\ref{Kee}). In the first case, this rule is provided by means of finding an
exponential ${\rm E}_\lambda(x)$ being dual to the initial one ${\rm
e}_\lambda(x)$ (this case is implemented for the basic- and $h$-calculi, and
the latter relates to the self-dual exponential); the second class requires to
deform the sum standing in the exponent of r.h.s. of Eq. (\ref{def}) according
to rules of the types (\ref{Keed}) and (\ref{Teed}). Let us consider above
cases separately.

Following to the method developed in the end of Section \ref{Sec.2},
generalization of both basic-deformed and $h$-sta\-tis\-tics is carried out in
the straightforward manner. However, as shows comparison of the considerations
stated in Sections \ref{Sec.3} and \ref{Sec.4}, to ensure the passage to the
expression
\begin{equation}
 {\rm Z}_{\lambda}\{J\}=\int{\rm e}_{\lambda}\left(-S\{\phi\}+J\cdot\phi\right)\{{\rm d}_\lambda\phi\}
 \label{Z(J)l1}
\end{equation}
that contains the ordinary sum of exponents one follows to use the deformed
Laplace transform
\begin{equation}
 {\rm Z}_{\lambda}\{J\}:=\int{\rm Z}_\lambda\{\phi\}{\rm E}_{\lambda}\{J\cdot\phi\}\{{\rm d}_\lambda\phi\}
 \label{Z(J)l2}
\end{equation}
with the dual exponential ${\rm E}_{\lambda}(x)\Rightarrow E_q(x)$, being
inherent in the basic-deformed calculus with $q=\lambda$, or the expression
\begin{equation}
 {\rm Z}_{\lambda}\{J\}:=\int{\rm Z}_\lambda\{\phi\}{\rm e}_{\lambda}\{J\cdot\phi\}\{{\rm d}_\lambda\phi\}
 \label{Z(J)l3}
\end{equation}
with the initial exponential ${\rm e}_{\lambda}(x)\Rightarrow e_h(x)$, taking
place in the $h$-deformed calculus with $h=\lambda$ [cf. Eqs. (\ref{Z(J)}) and
(\ref{Zh(J)})]. In all above cases, the $n$-fold derivative yields
\begin{eqnarray}
&&\left({\rm D}_{J_1}^\lambda\dots{\rm D}_{J_n}^\lambda\right) {\rm
Z}_\lambda\{J\}\\ &&=\int\big(\eta_\lambda
 \left(\phi_1\right)\dots\eta_\lambda
 \left(\phi_n\right)\big) {\rm
e}_{\lambda}\left(-S\{\phi\}+J\cdot\phi\right)\{{\rm d}_\lambda\phi\}.\nonumber
 \label{DDDl}
\end{eqnarray}
Here, we take into account the differentiation rule
\begin{equation}
 {\rm D}_{J_i}^{\lambda}{\rm e}_\lambda\left(-S+J\cdot\phi\right)=\eta_\lambda
 \left(\phi_i\right){\rm e}_\lambda(-S+J\cdot\phi)
  \label{lG}
\end{equation}
where an eigen-value $\eta_\lambda(\phi_i)$ is determined by action of a
generalized derivative ${\rm D}_{J_i}^{\lambda}$ with respect to a auxiliary
field $J_i$ (in the simple cases of both basic- and $h$-deformed calculi, one
has $\eta_\lambda(\phi_i)=\phi_i$, while the Tsallis calculus relates to
$\eta_\lambda(\phi_i)=\ln_{2-\lambda}(\phi_i)$ \cite{OSh}). As a result, the
correlator
\begin{eqnarray}
&&\left<\eta_\lambda
 \left(\phi_1\right)\dots\eta_\lambda
 \left(\phi_n\right)\right>_\lambda\\ \nonumber &&:={\rm
Z}_\lambda^{-1}\int\big(\eta_\lambda
 \left(\phi_1\right)\dots\eta_\lambda
 \left(\phi_n\right)\big) {\rm Z}_\lambda\{\phi\}\{{\rm d}_\lambda\phi\}
 \label{Cl}
\end{eqnarray}
with the partition function
\begin{equation}
 {\rm Z}_{\lambda}:=\int{\rm Z}_\lambda\{\phi\}\{{\rm d}_\lambda\phi\}
 =\int e_{\lambda}\left(-S\{\phi\}\right)\{{\rm d}_\lambda\phi\}
 \label{ZZl}
\end{equation}
is expressed in the form
\begin{eqnarray}
&&\left<\eta_\lambda
 \left(\phi_1\right)\dots\eta_\lambda
 \left(\phi_n\right)\right>_\lambda\nonumber\\ &&= {\rm Z}_\lambda^{-1}\left.\left({\rm
D}_{J_1}^\lambda\dots {\rm D}_{J_n}^\lambda\right) {\rm
Z}_\lambda\{J\}\right|_{J_1,\dots,J_n=0}.
 \label{CCl}
\end{eqnarray}

Within the harmonic approach, the action (\ref{S0}) is written in the
square-law form
\begin{equation}
S^{(0)}=\sum\limits_{i=1}^N\frac{(\phi_i)_\lambda^2}{[2]_\lambda}
 \label{Zzgg}
\end{equation}
where a field set $\{\phi_i\}$ is distributed with the unit variance and
$\lambda$-de\-for\-med square $(\phi_i)_\lambda^2$ and number $[2]_\lambda$ are
used instead of Eqs. (\ref{bn}) and (\ref{binom}). Then, the generating
functional (\ref{Z(J)l2}) takes the form
\begin{equation}
{\rm Z}_\lambda^{(0)}\{J\}=\prod_{i'}^{[N/2]}{\rm
z}_\lambda^{(0)}(J_{i'})\prod_{i{''}}^{[N/2]}{\rm
z}_{1/\lambda}^{(0)}(J_{i{''}})
 \label{Zzl1}
\end{equation}
of the type (\ref{Zz}). Respectively, for the generating functional
(\ref{Z(J)l3}) one obtains
\begin{equation}
{\rm Z}_\lambda^{(0)}\{J\}=\prod_i^N{\rm z}_\lambda^{(0)}(J_i).
 \label{Zzl2}
\end{equation}
Here, each of multipliers related to one site is determined by the expression
\begin{equation}
{\rm z}_\lambda^{(0)}(J)=\sqrt{[2]_\lambda\pi_\lambda}{\rm
E}_\lambda\left(\frac{ J^2}{2\Delta_\lambda^2}\right)
 \label{z0l}
\end{equation}
where $\Delta_\lambda$ is a $\lambda$-deformed variance taking the value
$\Delta_{\lambda_0}=1$ in the non-deformed limit $\lambda\to\lambda_0$; in
turn, $\lambda$-deformed $\pi$-number is defined by equation of the type
(\ref{I}) with $\lambda$ standing instead of $q$ (as pointed out above, a dual
$\lambda$-exponential ${\rm E}_\lambda(x)$ should be used in the case of the
type $q$-calculus and a self-dual $\lambda$-exponential ${\rm e}_\lambda(x)$ --
for the class of $h$-type calculus). By this, the specific partition function
${\rm z}_\lambda^{(0)}={\rm z}_\lambda^{(0)}(J=0)$ is given by the simple
expression
\begin{equation}
{\rm z}_\lambda^{(0)}=\sqrt{[2]_\lambda\pi_\lambda}.
 \label{G0l}
\end{equation}
Similarly to the $q$- and $h$-statistical field theories, the mean value of
free fields equals
$\left<\eta(\phi)\right>_\lambda^{(0)}\propto\left.J\right|_{J=0}=0$, while the
variance $\left<\eta^2(\phi)\right>_\lambda^{(0)}$ appears to be monotonically
increasing function of the deformation parameter $\lambda$. Calculation of
explicit form of dependences (\ref{z0l}) and (\ref{G0l}), as well as the
variance $\left<\eta^2(\phi)\right>_\lambda^{(0)}$ of order parameter
$\eta=\eta(\phi)$ related to the derivation rule (\ref{lG}) needs to specify a
concrete form of the $\lambda$-deformed exponentials.

The perturbation theory is based on the equation type of (\ref{E2}) and
(\ref{PT}), as well as the diagram technics and the Wick theorem are built
similarly to above considered field-theoretical schemes. Moreover, the additive
generating functional related to the non-additive ones, (\ref{Z(J)l2}) and
(\ref{Z(J)l3}), is expressed by the Green functional ${\rm
G}_\lambda^{(0)}\{J\}$ in the form (\ref{G}). The passage from this functional,
dependent on an auxiliary field $J$, to the conjugate functional
$\Gamma_\lambda=\Gamma_\lambda\{\phi\}$, being a functional of an order
parameter $\phi$, is achieved by the Legendre transformation type of (\ref{L}),
while the state equations have the form (\ref{SE}). Respectively, series of
sort (\ref{DJ}) and (\ref{DS}) have kernels that reduce to the $n$-particle
Green function and its irreducible part. Within the diagram representation,
these kernels look like depicted in Figure \ref{G.pdf}. What about the formal
equations for the generating functionals (\ref{Z(J)l2}) and (\ref{Z(J)l3}),
they take the forms (\ref{EM2}), (\ref{I2}), and (\ref{EM3}) to display a
system symmetry with respect to the $\lambda$-deformed variation
$\delta_\lambda\phi_i(x)$ and take into account an arbitrary condition
$F_j\{\phi\}=0$, $j=1,2,\dots$ for the set of fields $\{\phi_i\}$,
$i=1,2,\dots,N$. As pointed out above, in all above expressions the index $q$
should be substituted by the subscript $\lambda$, and the
$\lambda$-exponentials ${\rm e}_\lambda(x)$ and ${\rm E}_\lambda(x)$ should be
used instead of the $q$-exponentials $e_q(x)$ and $E_q(x)$.

Finally, we state main principles of generalization of the Kaniadakis-type
statistics. As was stressed in the beginning of this Section, the corner stone
of the Kaniadakis calculus is that the multiplication rule (\ref{Kee}) is
defined by deformed sum of the type (\ref{Keed}). As a result, the generating
functional (\ref{ZK(J)}) takes the form
\begin{eqnarray}
{\rm Z}_{\lambda}\{J\}&& :=\int{\rm Z}_\lambda\{\phi\}{\rm
e}_\lambda\{J\cdot\phi\}\{{\rm d}_\lambda\phi\}\nonumber\\ && =\int{\rm
e}_{\lambda}\left(-S\{\phi\}\stackrel{\lambda}{\oplus}J\cdot\phi\right)\{{\rm
 d}_\lambda\phi\}
 \label{ZG(J)}
\end{eqnarray}
determined by a generally deformed sum standing in the last exponent.
Respectively, the correlator (\ref{CCK}) is written as follows:
\begin{eqnarray}
&&\left<\eta_\lambda\left(\phi_{i_1}\right)\dots\eta_\lambda
\left(\phi_{i_n}\right)\right>_\lambda\nonumber\\ &&= {\rm
Z}^{-1}_\lambda\left.\left({\rm D}_{J_1}^{\lambda\left(\phi_1\right)}\dots{\rm
D}_{J_n}^{\lambda\left(\phi_n\right)}\right){\rm
Z}_\lambda\{J\}\right|_{J_1,\dots,J_n=0}
 \label{CCG}
\end{eqnarray}
with the partition function ${\rm Z}_\lambda={\rm Z}_\lambda\{J=0\}$. Instead
of Eq. (\ref{lG}), we take into account here the differentiation rule
\begin{equation}
{\rm D}_x^{\lambda(a)}{\rm e}_\lambda(ax)=\eta_\lambda(a){\rm e}_\lambda(ax)
\label{ruled}
\end{equation}
with an eigen-value $\eta_\lambda(a)$ and a deformation $\lambda(a)$, being
defined by action of a generalized derivative ${\rm D}_x^{\lambda(a)}$ with
respect to a variable $x$ at arbitrary constant $a$ (for the Kaniadakis
calculus, one has $\eta_\lambda(a)=a$ and $\lambda(a)=\lambda a$).

By analogy with above considered field-theoretical sche\-mes, one finds
partition function, perturbation theory, diagram technics, additive and
conjugated Green functional, many-particle Green functions and their
irreducible parts, as well as formal equations for the generating functional of
systems that display a symmetry with respect to field variation and have some
constraints.

\section{Concluding remarks}\label{Sec.7}

Before a discussion of the results obtained we should stress uppermost that
quantum algebra and quantum groups, on whose basis our approach is founded,
have been the subject of intense research in different fields of the
$q$-deformed quantum theory (see \cite{Wilczek} and references therein).
Moreover, using the $q$-deformed algebra has allowed to develop multifractal
theory \cite{OlKhB,23} and thermostatistics of deformed bosons and fermions
\cite{21}. In our consideration, we restrict ourselves with generalization of
the only classical thermostatistics, whose version has been developed first in
work \cite{BasicDef}. The principle peculiarity of the basic-deformed
distribution arising in this model is to exhibit a cut-off in the energy
spectrum which is generally expected in complex systems, whose underlying
dynamics is governed by long-range interactions.

It is worthwhile to note in this connection the deformed distribution proposed
by Tsallis \cite{4} that is characterized by the power-law asymptotic behavior.
The Tsallis picture is known to be inherent in self-similar statistical
systems, whose field theory has been built by using both Mellin transform of
the Tsallis exponential and Jackson derivative \cite{OSh}. Contrary to the
examples considered in Sections \ref{Sec.3}--\ref{Sec.6}, a fluctuating order
parameter of self-similar systems has non-zero mean value that reduces to the
Tsallis deformed logarithm of the amplitude of a hydrodynamic mode. Formally,
this is caused by using the Mellin transform with the power-law kernel
$\phi^J=\exp[J\ln(\phi)]$ instead of the Fourier-Laplace one with the
exponential kernel $\exp(J\phi)$.

Above, we develop field-theoretical schemes founded on the basis of
basic-deformed and finite-difference calculi \cite{Kac}, as well as within
framework of deformation procedures proposed by Tsallis, Abe, Kaniadakis, and
Naudts \cite{4,6,7,Naudts}. We construct generating functionals related and
find their connection with corresponding correlators for basic-deformed, $h$-,
and Kaniadakis calculi. Moreover, we introduce pair of additive functionals
whose expansions into deformed series yield both Green functions and their
irreducible proper vertices, as well as find formal equations, governing by the
generating functionals of systems which possess a symmetry with respect to a
field variation and are subjected to an arbitrary constrain. Finally, we
generalize in the Naudts spirit the field-theoretical schemes inherent in
concrete calculi.

Concerning the physical results above obtained, one should point out
peculiarities of dependences of both one-site partition function and variance
of free fields on deformations (see Figures \ref{z0q} and \ref{z0hh},
respectively). In the case of basic deformation, the specific partition
function has in logarithmic axes symmetrical form with respect to the maximum
point $q=1$ (by this, the $\Delta^2$ increase shifts up the curves of
dependences related logarithmically equidistantly). For the $h$-deformation,
the specific partition function takes non-deformed value. In the case of the
Kaniadakis deformation, the dependence of the one-site partition function
$z_\kappa^{(0)}$ on the deformation parameter $\kappa$ has a symmetrical form
with respect to the point $\kappa=0$ (by this, the $\kappa$ growth results in
the $z_\kappa^{(0)}$ increase at small inverse curvature $\Delta^2$ of the
effective action, while the $\Delta$ growth transforms the concave curve of the
dependence $z_\kappa^{(0)}(\kappa)$ into the convex one). What about the
correlators of free-distributed fields, the only moment of the second order
takes non-zero values. For all distributions, this moment is proportional to
the inverse curvature $\Delta^2$ of the action related to increase with the
deformation parameter growth linearly in the case of the basic-deformed
statistics and non-linearly rapidly for the $h$-statistics. More complicated
behaviour takes place for the Kaniadakis deformation, when the variance related
decreases first and increases then up to an anomalous value.

\begin{acknowledgement}
The authors are very grateful to the organizing committee of the
NEXT-$\Sigma\Phi$ 2008 conference in Kolymbari, Crete, Greece. We thank
especially G. Kaniadakis and A. Scarfone for sending us the special issue of
the European Physical Journal containing several papers reported in this
conference.
\end{acknowledgement}

\section*{Appendix A}\label{Sec.8}
 \def\theequation{{A}.\arabic{equation}}
 \setcounter{equation}{0}

Within framework of the basic-deformed calculus, the pair of dual $q$-gamma
functions is defined by the integrals
\begin{eqnarray}
 \gamma_q(\alpha):=\int\limits_{0}^{+\infty}x_q^{\alpha-1}e_q(-x){\rm
 d}_qx,\label{gamma}\\
 \Gamma_q(\alpha):=\int\limits_{0}^{+\infty}x_q^{\alpha-1}E_q(-qx){\rm d}_qx
  \label{Gamma}
\end{eqnarray}
where the upper limit equals $\frac{1}{1-q}$ for $|q|<1$. These definitions are
principle different from the introduced in Ref. \cite{Kac} because we
substitute the ordinary power function $x^{\alpha-1}$ by the deformed one
$x_q^{\alpha-1}$ which generalizes the deformed binom (\ref{binom}) to
substitute integer $n$ by arbitrary exponent $\alpha-1$. The definitions
(\ref{gamma}) and (\ref{Gamma}) ensure the properties
\begin{eqnarray}
&&\Gamma_q(\alpha+1)=[\alpha]_q\Gamma_q(\alpha),\\
&&\gamma_q(\alpha+1)=[\alpha]_q\gamma_q(\alpha)q^{-\alpha}=-[-\alpha]_q\gamma_q(\alpha)
  \label{gq1p}
\end{eqnarray}
with $\gamma_q(0)=\gamma_q(1)=1$ and $\Gamma_q(0)=\Gamma_q(1)=1$. At
$\alpha=1/2$, the definition (\ref{gamma}) gives the deformed Poisson integral
\begin{equation}
\int\limits_{-\infty}^{+\infty}e_q\big(-x_q^2\big)d_qx=\frac{2}{[2]_q}\gamma_q(1/2)\equiv\sqrt{\pi_q}
  \label{I}
\end{equation}
where deformed $\pi$-number $\pi_q$ has the limit $\pi_{q\to 1}=\pi\equiv
3.14159\dots$

The specific generating functional is calculated as:
\begin{eqnarray}
&&
z_q^{(0)}(J)=\int\limits_{-\infty}^{+\infty}e_q\left(-\frac{\phi^2}{[2]_q\Delta^2}\right)
E_{q}(J\phi){\rm d}_q\phi\nonumber\\ && =E_q\left(q\frac{\Delta^2
J^2}{[2]_q}\right)\int\limits_{-\infty}^{+\infty}e_q
\left\{-\frac{\left(\phi-\Delta^2J\right)_q^2 }{[2]_q\Delta^2}\right\}{\rm
d}_q\phi\nonumber\\ && =E_q\left(q\frac{\Delta^2
J^2}{[2]_q}\right)\int\limits_{-\infty}^{+\infty}e_q \left(-\frac{x_q^2
}{[2]_q\Delta^2}\right){\rm d}_qx\nonumber\\ &&
=\frac{2\Delta}{\sqrt{[2]_q}}\gamma_q\left(\frac{1}{2}\right)E_q\left(q\frac{\Delta^2
J^2}{[2]_q}\right)
  \label{zJq}
\end{eqnarray}
where the variable $x=\phi-\Delta^2J$ is introduced.

In the case of the $h$-deformed calculus, the gamma function related
\begin{equation}
 \gamma_h(\alpha):=\int\limits_{0}^{+\infty}x^{\alpha-1}e_h(-x){\rm
 d}_hx=[1]_h^{\alpha-1}\Gamma(\alpha)\label{gammah}
\end{equation}
appears to be proportional to the ordinary one $\Gamma(\alpha)$. By analogy
with the calculations (\ref{zJq}), the one-site generating functional is
expressed as:
\begin{eqnarray}
&&z_h^{(0)}(J)=2\frac{[1]_h}{\sqrt{[2]_h}}\gamma_h\left(\frac{1}{2}\right)
e_h\left[\frac{[1]_h}{2}(\Delta J)^2\right]\Delta \nonumber\\ &&=
\sqrt{[2]_h}\gamma_h\left(\frac{1}{2}\right) e_h\left[\frac{[1]_h}{2}(\Delta
J)^2\right]\Delta.
  \label{zJh}
\end{eqnarray}
Taking into account the property (\ref{gammah}), one obtains the simple result
(\ref{z01}).


\begin{thebibliography}{}
\bibitem{4} C. Tsallis, J. Stat. Phys. {\bf 52}, 479 (1988)
\bibitem{5} E.M. Curado, C. Tsallis, J. Phys. A {\bf 24}, L69 (1991)
\bibitem{TMP} C. Tsallis, R.S. Mendes, A.R. Plastino, Physica A {\bf 261}, 534 (1998)
\bibitem{6} S. Abe, Phys. Lett. A {\bf 224}, 326 (1997)
\bibitem{7} G. Kaniadakis, Phys. Rev. E {\bf 66}, 056125 (2002)
\bibitem {Naudts} J. Naudts, Physica A  {\bf 340}, 32 (2004)
\bibitem{8} G. Kaniadakis, M. Lissia, A.M. Scarfone, Physica A {\bf 340}, 41
(2004); Phys. Rev. E {\bf 71}, 046128 (2005)
\bibitem{9} A.M. Scarfone, T. Wada, Phys. Rev. E {\bf 72}, 026123 (2005)
\bibitem{10} A. Lavagno, A.M. Scarfone, N.P. Swamy, Eur. Phys. J. B {\bf 50}, 351 (2006)
\bibitem{BasicDef} A. Lavagno, A. M. Scarfone, P. Narayana Swamy,
J. Phys. A {\bf 40}, 8635 (2007)
\bibitem{1} R. Hilborn, {\it Chaos and Nonlinear Dynamics: An Introduction for Scientists and
Engineers} (Oxford University Press, Oxford, 2001)
\bibitem{2} C. Beck, F. Schl\"ogl, {\it Thermodynamics of chaotic
system} (Cambridge University Press, Cambridge, 1993)
\bibitem{3} S. Abe, {\it Nonextensive Statistical Mechanics and its Applications}, edited by Y. Okamoto (Springer, Berlin, 2001)
\bibitem{T} C.~Tsallis, {\it Introduction to Nonextensive Statistical Mechanics -- Approaching a Complex World} (Springer, New York, 2009)
\bibitem{Callen} H.B. Callen, {\it Thermodynamics and an Introduction to Thermostatistics} (Wiley, New York, 1985)
\bibitem{OSh} A. Olemskoi, I. Shuda, Phys. Lett. A {\bf 373}, 4012 (2009)
\bibitem{Par} G. Parisi, {\it Statistical Field Theory} (Addison-Wesley, Redwood City, 1988)
\bibitem{Zinn} J. Zinn-Justin, {\it Quantum Field Theory and Critical
Phenomena} (Clarendon Press, Oxford, 1993)
\bibitem{15} E. Heine, J. Reine Angew. Math. {\bf 32}, 210 (1846);
J. Reine Angew. Math. {\bf 34}, 285 (1847)
\bibitem{16} F.H. Jackson, Am. J. Math. {\bf 38}, 26 (1909);
Mess. Math. {\bf 38}, 57 (1909)
\bibitem{17} G. Gasper, M. Rahman, {\it Basic hypergeometric series, Encyclopedia of mathematics and its applications} (Cambridge Univeristy Press, Cambridge, 1990)
\bibitem{18} H. Exton, {\it q-Hypergeometric functions and applications} (Ellis Horwood, Chichester, 1983)
\bibitem{19} E. Celeghini et al., Ann. Phys. {\bf 241}, 50 (1995)
\bibitem{20} R.J. Finkelstein Int. J. Mod. Phys. A {\bf 13}, 1795 (1998)
\bibitem{21} A. Lavagno, P. Narayana Swamy, arXiv:0912.4596
\bibitem{22} A. Erzan, J.-P. Eckmann, Phys. Rev. Lett. {\bf 78}, 3245 (1997)
\bibitem{23} A.I. Olemskoi, I.A. Shuda, V.N. Borisyuk, Europhys. Lett.  {\bf 89}, 50007 (2010)
\bibitem{24} D. Sornette, {\it Critical phenomena in natural
Sciences} (Springer, New York, 2001)
\bibitem{25} D. Sornette, Phys. Rep. {\bf 297}, 239 (1998);
W.-X. Zhou, D. Sornette, Phys. Rev. E {\bf 66}, 046111 (2002);
arXiv:cond-mat/0408600v2
\bibitem{26} S. Gluzman, D. Sornette, Phys. Rev. E {\bf 65} 036142 (2002)
\bibitem{OVSh} A.I. Olemskoi, A.S. Vaylenko, I.A. Shuda, Physica A {\bf 388},
1929 (2009)
\bibitem{Kac} V. Kac, P. Cheung, {\it Quantum Calculus} (Springer-Verlag, New York, 2002)
\bibitem{Ks} G. Kaniadakis, Eur. Phys. J. A {\bf 40}, 275 (2009)
\bibitem{Kan} G. Kaniadakis, Eur. Phys. J. B {\bf 70}, 3 (2009)
\bibitem{M} D.P. Mittal, Metrika {\bf 22}, 35 (1975)
\bibitem{ST} B.D. Sharma, I.J. Taneja, Metrika {\bf 22}, 205 (1975)
\bibitem{Borges} E.P. Borges, Physica A {\bf 340}, 95 (2004)
\bibitem{Wilczek} F. Wilczek, {\it Fractional Statistics and Anyon Superconductivity} (World
Scientific, Singapore, 1990)
\bibitem{OlKhB} A.I. Olemskoi, V.O. Kharchenko, V.N. Borisyuk, Physica A
{\bf 387}, 1895 (2008)
\end{thebibliography}
\end{document}